\documentclass[floatfix,lengthcheck,showpacs,amssymb,amsmath,amsfonts,twocolumn,nolinenumber]{aastex63}
\usepackage{CJK}
\usepackage[utf8]{inputenc}
\usepackage{color}
\usepackage{graphicx}
\usepackage{float}
\usepackage{subfigure}
\usepackage{amsmath}
\usepackage{afterpage}
\usepackage{acro}
\usepackage{multirow}
\usepackage{import}
\usepackage{svn-multi}
\usepackage{hyperref}
\usepackage{longtable}
\usepackage{xcolor,colortbl}
\definecolor{Gray}{gray}{0.9}
\usepackage[left]{lineno}

\DeclareAcronym{GR}{
	short = GR,
	long  = general relativity
	}
	
\DeclareAcronym{BH}{
	short = BH ,
	long  = black hole
}

\DeclareAcronym{BBH}{
	short = BBH ,
	long  = binary black hole
}

\DeclareAcronym{NSBH}{
	short = NSBH ,
	long  = neutron star black hole
}

\DeclareAcronym{GW}{
	short = GW ,
	long  = gravitational wave
}

\DeclareAcronym{CBC}{
	short = CBC,
	long  = compact binary coalescence
}

\DeclareAcronym{SNR}{
	short = SNR,
	long  = signal-to-noise ratio
}

\DeclareAcronym{GRB}{
	short = GRB,
	long  = gamma-ray burst
}

\DeclareAcronym{FAR}{
	short = FAR,
	long  = false alarm rate
}

\DeclareAcronym{EM}{
	short = EM,
	long  = electromagnetic
}

\usepackage[capitalise]{cleveref}
\crefname{figure}{Fig.}{Figs.}
\Crefname{figure}{Fig.}{Figs.}

\def\be{\begin{equation}}
\def\ee{\end{equation}}
\def\({\left(}
\def\){\right)}
\def\[{\left[}
\def\]{\right]}

\def\msun{$M_\odot$}
\begin{document}
\begin{CJK*}{UTF8}{gbsn}
\title{Search for Coincident Gravitational Wave and Long Gamma-Ray Bursts from 4-OGC and the Fermi-GBM/Swift-BAT Catalog}
\correspondingauthor{Yi-Fan Wang (王一帆)}
\email{yifan.wang@aei.mpg.de}

\author[0000-0002-2928-2916]{Yi-Fan Wang (王一帆)}
\author[0000-0002-1850-4587]{Alexander H. Nitz}
\affiliation{Max-Planck-Institut f{\"u}r Gravitationsphysik (Albert-Einstein-Institut), D-30167 Hannover, Germany}
\affiliation{Leibniz Universit{\"a}t Hannover, D-30167 Hannover, Germany}

\author[0000-0002-0355-5998]{Collin D. Capano}
\affil{Department of Mathematics, University of Massachusetts, Dartmouth, MA 02747, USA}
\affil{Max-Planck-Institut f{\"u}r Gravitationsphysik (Albert-Einstein-Institut), D-30167 Hannover, Germany}
\affil{Leibniz Universit{\"a}t Hannover, D-30167 Hannover, Germany}

\author[0000-0002-9738-1238]{Xiangyu Ivy Wang}
\affiliation{School of Astronomy and Space Science, Nanjing
University, Nanjing 210093, China}
\affiliation{Key Laboratory of Modern Astronomy and Astrophysics (Nanjing University), Ministry of Education, China}

\author[0000-0003-0691-6688]{Yu-Han Yang}
\affiliation{School of Astronomy and Space Science, Nanjing University, Nanjing 210093, China}
\affiliation{Key Laboratory of Modern Astronomy and Astrophysics (Nanjing University), Ministry of Education, China}

\author[0000-0003-4111-5958]{Bin-Bin Zhang}
\affil{School of Astronomy and Space Science, Nanjing
University, Nanjing 210093, China}
\affil{Key Laboratory of Modern Astronomy and Astrophysics (Nanjing University), Ministry of Education, China}


\begin{abstract}
The recent discovery of a kilonova associated with an apparent long-duration gamma-ray burst has challenged the typical classification that long gamma-ray bursts originate from the core collapse of massive stars and short gamma-ray bursts are from compact binary coalescence.
The kilonova indicates a neutron star merger origin and suggests the viability of gravitational-wave and long gamma-ray burst multimessenger astronomy.
Gravitational waves play a crucial role by providing independent information for the source properties.
This work revisits the archival 2015-2020 LIGO/Virgo gravitational-wave candidates from the 4-OGC catalog which are consistent with a binary neutron star or neutron star-black hole merger and the long-duration gamma-ray bursts from the {Fermi-GBM} and {Swift-BAT} catalogs.
We search for spatial and temporal coincidence with up to 10 s time lag between gravitational-wave candidates and the onset of long-duration gamma-ray bursts.
The most significant candidate association has only a false alarm rate of once every two years; given the LIGO/Virgo observational period, this is consistent with a null result.  
We report an exclusion distance for each search candidate for a fiducial gravitational-wave signal with conservative viewing angle assumptions. 
\end{abstract}

\acresetall
\section{Introduction} \label{sec:intro}
The {direct} detection of a \ac{GW} from the binary neutron star merger GW170817 together with an electromagnetic counterpart began a new era of multi-messenger astronomy \citep{TheLIGOScientific:2017qsa,Monitor:2017mdv,GBM:2017lvd}. The initial report of the \ac{GW} by Advanced LIGO \citep{TheLIGOScientific:2014jea} and Virgo \citep{TheVirgo:2014hva} and the short \ac{GRB} GRB 170817A \citep{2017ApJ...848L..14G} by the Fermi Gamma-ray Burst Monitor (GBM) \citep{2009ApJ...702..791M} triggered an extensive campaign of follow-up observations across the \ac{EM} spectrum \citep[e.g.,][]{2017ApJ...848L..17C,2017ApJ...848L..15S,2017ApJ...848L..25H}. 
This multimessenger event provided a great wealth of knowledge for astrophysics and fundamental physics. For example, the speed of \ac{GW} propagation was constrained to deviate by no more than a factor of $[-3 \times10^{-15}, 7\times10^{-16}]$ from the speed of light \citep{Monitor:2017mdv,LIGOScientific:2018dkp}. Besides, direct measurement of the Hubble constant was found to be $H_0=70.0^{+12.0}_{-8.0}$ km s$^{-1}$ Mpc$^{-1}$ within the 68$\%$ credible interval \citep{LIGOScientific:2017adf} \citep[see also e.g.,][for later improvements]{Guidorzi:2017ogy, Hotokezaka:2018dfi}.
It also confirmed the {long-held} hypothesis that short-duration \ac{GRB}s originate from the merger of compact objects involving neutron stars \citep{1986ApJ...308L..47G,1986ApJ...308L..43P,cite-key,1992ApJ...395L..83N}, and provided the first confirmation for a kilonova associated with a neutron star merger and r-process nucleosynthesis  \citep[e.g.,][]{LIGOScientific:2017pwl}. 

The correlation analysis of signals from different observations underlies \ac{GW}/\ac{EM} multimessenger astronomy.
Since the beginning of the third observational run in April 2019, LIGO/Virgo have released real-time public trigger alerts \citep{LIGOScientific:2020ibl}, enabling swift follow-up observation of \ac{EM}-band or neutrino signals.
Later, archival GW/EM searches are performed to dig deeper for associated signals missed by the low-latency searches.
For instance, \cite{LIGOScientific:2016akj,LIGOScientific:2019obb,LIGOScientific:2020lst,LIGOScientific:2021iyk} and \cite{Nitz:2019bxt} have studied temporal and spatial coincidence between \ac{GW} and \ac{GRB} using archival LIGO/Virgo data and {Fermi-GBM/Swift-Burst Alert Telescope (BAT)} \citep{2009ApJ...702..791M,SwiftScience:2004ykd,2005SSRv..120..143B} data. \cite{2022arXiv220312038T} and \cite{Wang:2022ryc} have searched for GWs coincident with observations from the fast radio burst catalog released by the CHIME collaboration \citep{CHIMEFRB:2018mlh,CHIMEFRB:2021srp}.

Conventionally, \acp{GRB} are separated into two classes based on their duration and spectral characteristics, i.e., short-duration hard-spectral and long-duration soft-spectral \citep{1992AIPC..265..304D,1993ApJ...413L.101K}.
As confirmed by GW170817, short GRBs are known to originate from neutron star mergers, while many long GRBs are generated by the core collapse of massive stars as determined by the co-observation of supernova \citep{1998Natur.395..670G,2003Natur.423..847H,2003ApJ...591L..17S}. 
Based on this knowledge, previous searches \citep{LIGOScientific:2016akj,LIGOScientific:2019obb,LIGOScientific:2020lst,LIGOScientific:2021iyk,Nitz:2019bxt} for GW/GRB {coincidences} have only targeted short GRBs with a template-based match-filtering method to look for associated \ac{GW} signals from compact binary coalescence. The \ac{GW} candidates potentially associated with long GRBs were analyzed with a template-free generic search, with the aim to detect \acp{GW} from asymmetric core-collapse massive stars.

However, the separation of GRBs based on their observed light-curve duration is not precise~\citep{Gehrels:2006tk,Gal-Yam:2006qle,Zhang:2021agu,Ahumada:2021zrl}. In particular, \cite{Rastinejad:2022zbg} recently reported a kilonova located at 350 Mpc associated with GRB 211211A, an apparent long-duration GRB with $T_{90}=51.37$s given by \cite{GRB211211A}, challenging the previous paradigm for GRB classification and progenitor.
Additional interesting features were reported, including a quasi-periodic oscillated precursor that occurred $\sim1$ s prior to the main emission \citep{Xiao:2022quv}, and a high energy ($>100$ MeV) gamma-ray afterglow lasting $\sim2\times10^4$ s that occurred $\sim1000$ s after the burst \citep{Mei:2022ncd,Zhang:2022fzj}.
Several models have been proposed to explain the emission mechanism of GRB 211211A; for instance, fast-cooling synchrotron radiation after a binary neutron star merger \citep{Gompertz:2022jsg}, magnetic barrier effect involving a magnetar progenitor \citep{Gao:2022iwn}, a neutron star-white dwarf merger \citep{Yang:2022qmy}, or thermal emission from heated dust as an alternative scenario to kilonova \citep{Waxman:2022pfm}.
The associated kilonova clearly indicates the progenitor of GRB 211211A is a binary merger with at least one neutron star, suggesting the prospect of detecting a \ac{GW} signal from events of this kind \citep[however][for an alternative explanation]{Waxman:2022pfm}.
\ac{GW} observation provides unique information about the source's mass and spin~\citep{Veitch:2014wba}, which may help disentangle different models~\citep{Gompertz:2022jsg,Gao:2022iwn,Yang:2022qmy,Waxman:2022pfm}.
Unfortunately, none of the \ac{GW} detectors were in observation mode at the time of the GRB 211211A.

Nevertheless, we revisit the archival GW data from LIGO/Virgo and long GRB candidates from the {Fermi-GBM} and {Swift-BAT} observation motivated by the kilonova/GRB 211211A association.
We focus on \ac{GW} candidates, both significant and subthreshold, that are consistent with mergers involving at least one neutron star and study their temporal and spatial coincidence with confident long GRB observations recorded by {Fermi-GBM} and {Swift-BAT}.

\section{Candidate selection and ranking statistic}\label{sec:can}

This section describes the selection of GW and long GRB candidates, and our algorithm to rank the GW/long-GRB temporal and spatial coincidence.

Advanced LIGO and Virgo have completed three observational runs from 2015 to 2020 and released around ninety GW events in the Gravitational Wave Transient Catalog \citep{LIGOScientific:2021djp}. Additional detections were reported by independent groups \citep{Nitz:2021zwj, Olsen:2022pin}.
This work utilizes the GW search results from the fourth Open Gravitational-wave Catalog \citep[4-OGC, ][]{Nitz:2021zwj}.
4-OGC has searched for the entire three observation runs of Advanced LIGO/Virgo using the \texttt{PyCBC} toolkit \citep{pycbc-github} and detected ninety-four confident \ac{GW} events from compact binary coalescences. The subthreshold triggers' search results are also publicly released \citep{4-OGC}.

To select potential GW signals comprising at least one neutron star, we choose those confident and subthreshold candidates in 4-OGC where the secondary mass $m_2$ (the mass of the lighter component of the binary) of the associated gravitational-wave template is within [1,3] \msun.
Given the observation of low-spin neutron star -- black hole mergers~\citep{LIGOScientific:2021qlt}, we also extend the mass range up to total mass $m_1+m_2$ of 10 \msun{}. 
Since we primarily target EM counterpart from binary neutron star mergers, we constrain the effective spin $\chi_\mathrm{eff}$ in $[-0.2, 0.2]$. 
Effective spin is the primary spin parameter characterizing the GW signal \citep{Ajith:2009bn} and is defined as
\be
\chi_\mathrm{eff} = \frac{m_1\chi_1 + m_2\chi_2 }{ m_1+m_2},
 \ee 
where $\chi_{1/2}$ is the dimensionless component spin aligned with the orbital angular momentum direction. 
We limit $\chi_\mathrm{eff}$ because observations show that galactic binary neutron stars would have small spin ($\chi_\mathrm{eff}\lesssim0.05$) at merger \citep{PhysRevD.98.043002}; the binary neutron star events GW170817 and GW190425 are consistent with zero spin \citep{LIGOScientific:2018hze, Abbott:2020uma}.
Also, note that [1,2] \msun{} is considered the binary neutron star search region in 4-OGC, and $\chi_\mathrm{eff}$ in the search template is constrained to [-0.05,0.05].
While we expect high-spin black hole -- neutron star mergers to experience greater disruption~\citep{Foucart:2018rjc, Capano:2019eae}, our candidate population may have significant observational biases if the spin-axis is misaligned with the orbital axis~\citep{Dhurkunde:2022aek}, which can be improved in future GW search by accounting for spin precession. 
We also expect that for neutron star -- black hole mergers with a mass ratio greater than $\sim$10:1 there is not likely to be sufficient mass ejecta to produce a GRB~\citep{Foucart:2018rjc, Capano:2019eae}.

We further require the candidates to trigger at least two GW detectors, i.e., no single detector triggers are considered \citep{Nitz:2020naa}.
{Single detector triggers have less precise sky localization, typically with uncertainty up to tens of thousands of square degrees. 
Two or three detector detections, on the other hand, can pinpoint the source to within tens of square degrees in the most precise cases, thereby significantly reducing background noise contamination.
Furthermore, quantitative estimates indicate that only $6\%$ of horizon volume-time will be lost during the first three observation runs of LIGO/Virgo if single detector observations are excluded.
Consequently, we do not consider single detector triggers in light of the aforementioned factors.
}
The 4-OGC catalog assigns a search ranking statistic, $\lambda_\mathrm{gw}$, to each GW trigger, which is the natural logarithm of the ratio of rate densities for signal and noise hypothesis \citep{Nitz:2017svb, Davies:2020tsx}.
We only consider those triggers with $\lambda_\mathrm{gw}\geq0$. These choices resulted in {$\sim 5\times 10^5$} triggers from the three observational runs in 2015-2022.

For each \ac{GW} candidate, we use \texttt{PyCBC Inference} \citep{Biwer:2018osg} to estimate the Bayesian posterior of the sky localization.
To save the computation resources, we fix the mass and spin parameters to be those reported by 4-OGC \citep{4-OGC} from the PyCBC search.
We do not expect this procedure would significantly bias the sky position estimation because of the decoupling of intrinsic parameters (the source parameters independent of the observer orientation) and extrinsic parameters (the source parameters dependent on the observer orientation) \citep{PhysRevD.93.024013}.
Therefore, the variables to be estimated include the luminosity distance, right ascension, declination, polarization angle, inclination angle between the source orbital angular momentum and the line of sight of observatories, and the phase and time of coalescence.
The posterior is numerically estimated using the dynamical nested sampler \textit{dynesty} \citep{speagle:2019} using the standard \ac{GW} likelihood~\cite{Finn:1992wt} assuming stationary and Gaussian noise.

For long GRBs, we select those from the Fermi-GBM and Swift-BAT, with $T_{90} - \delta T_{90} > $ 4s, where $T_{90}$ and $\delta T_{90}$ are the time duration containing $90\%$ of the burst fluence and its associated error, respectively.
The sky maps of these long-GRBs are released by Fermi-GBM and Swift-BAT, respectively~\citep{2009ApJ...702..791M,SwiftScience:2004ykd}. 

We describe the algorithm to quantify the temporal and spatial coincidence of GW and long GRB.
The time lag between GW signal and long GRB is allowed to be within [0,10] s.  
{We select the search window based on the expected delay time between the GW signal and the emission of the GRB. 
For GRB 170817A, such delay time, $\Delta t_{\rm GW-GRB}$, is $\sim 1.7$ s, which coincides with the burst duration of $T_{\rm 90}\sim $ 2 s \citep{2018NatCo...9..447Z}. The coincidence can be explained by a magnetized jet dissipating in an optically thin region in a large emission radius \citep{2018NatCo...9..447Z}. Assuming such a model can also be applied to the recently discovered merger-type long-duration GRB 211211A \citep{Yang:2022qmy}, we applied our search window to 10 seconds in accordance with the duration of the main peak of GRB 211211A.} 
GW/long GRB pairs which meet this condition are considered temporally associated.
We further compute the sky overlap probability for the associated pairs using the posterior overlap integral 
following Eq. 11 in \cite{Ashton:2017ykh} as
\be
\mathcal{B}_\mathrm{overlap} = \int \frac{P(\vec \theta|d_\mathrm{GW}, \mathcal{H})P(\vec \theta|d_\mathrm{GRB}, \mathcal{H})}{P(\vec \theta| \mathcal{H})} d \vec\theta,
\ee
where $\vec\theta$ is the sky localization (right ascension and declination), $P(\vec \theta|d_\mathrm{GW}, \mathcal{H})$ and $P(\vec \theta|d_\mathrm{GRB}, \mathcal{H})$ are the sky location posterior estimated by \ac{GW} and \ac{GRB} data, respectively, $P(\vec \theta| \mathcal{H})$ is the prior that we choose to be isotropic.  $\mathcal{H}$ is the underlying hypothesis characterizing the GW and GRB data analysis.

Given the search ranking statistic $\lambda_\mathrm{gw}$ from PyCBC, we design the following ranking statistic for \ac{GW}/long-\ac{GRB} association 
\begin{equation}
\lambda_\mathrm{gw+lgrb} = \lambda_\mathrm{gw} + \ln \mathcal{B}_\mathrm{overlap}.
\end{equation}

To measure the statistical significance for a specific associate GW/long-GRB pair, we time shift the GRB data with respect to GW data with a time stride larger than 10 s (we use 200 s) and calculate the ranking statistics of the new associated pairs.  Any coincidence obtained after the time-shifting is non-astrophysical and thus considered as the background distribution from the null hypothesis that GW and GRB candidates are only chance associations.
Performing the time-shifting multiple times, we effectively created a background observation of $\sim 1000$ years.
The false alarm rate for a foreground association is quantified by the rate of background signals with an equal or more significant ranking statistic.
Following the conventions in the literature \citep[e.g.,][]{TheLIGOScientific:2016pea}, we consider \ac{GW} candidate events that have a false alarm rate less than 1 in 100 years to be a confident detection.

\section{Search results and implications}

\begin{figure}[htp]
    \centering
\includegraphics[width=\columnwidth]{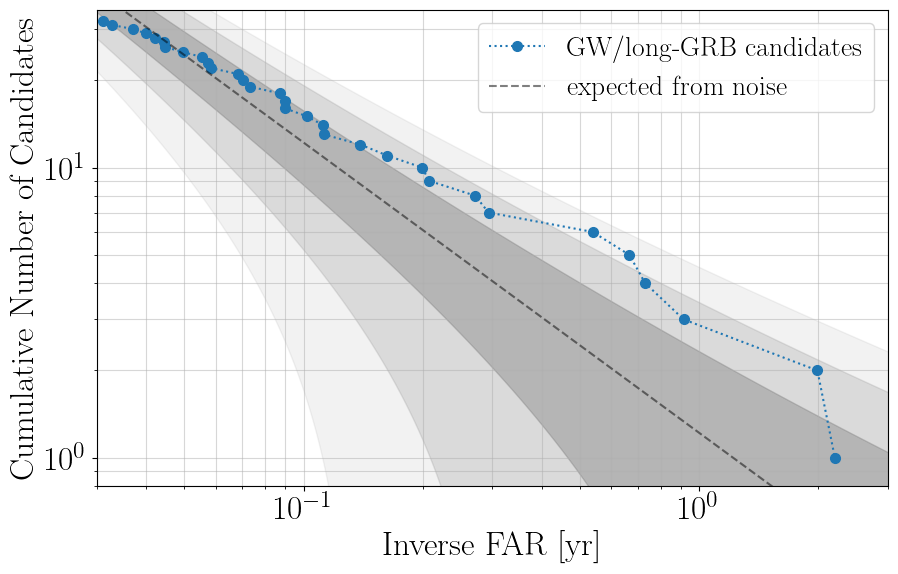}
    \caption{Cumulative number of GW/long-GRB association candidates versus the inverse false alarm rate. 
    The dashed line and shadows show the expected results from background noise fluctuation and one-, two- and three-sigma uncertainty from a Poisson process.
    The most significant associated pair $200205\_201716$ GRB 200205845 has ranking statistic $\lambda_\mathrm{gw+grb}$=5.53 which corresponds to a false alarm rate of 1 in 2.2 years.}
    
    \label{fig1}
\end{figure}

As described, we target the long GRB signals that occurred at most 10 s after a GW candidate and rank their spatial overlap and statistical significance.
The complete search results are illustrated in \cref{fig1} for the cumulative number of candidates and the inverse false alarm rate from search and expected from noise background, and \cref{table} for more detailed source information for the associated pairs.

\begin{figure}[htp]
    \centering
    \includegraphics[width=\columnwidth]{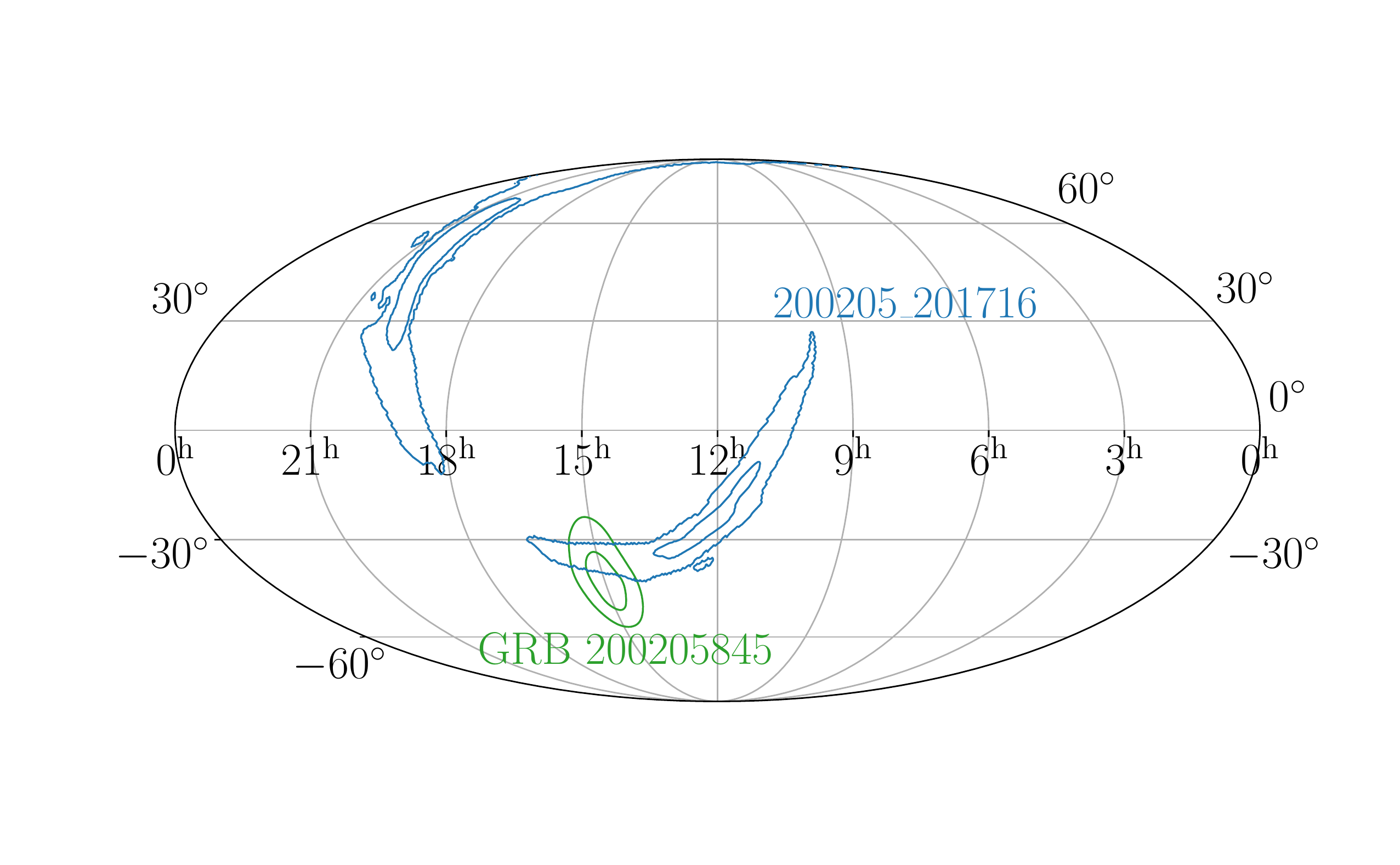}
    \caption{The posterior distribution of sky location for $200205\_201716$ from Bayesian inference and GRB 200205845 from the Fermi-GBM burst catalog.
        The inner and outer contours correspond to 50$\%$ and $90\%$ credible intervals, respectively. 
    The sky map overlap integral is $\ln B_\mathrm{overlap}=1.35$.
    }
    \label{fig:sky}
\end{figure}

There are thirty-two associated GW/long-GRB candidates from 2015-2020 when the GW and GRB detectors were both in observation mode.
The most significant candidate is from the \ac{GW} candidate $200205\_201716$ and the long-duration GRB 200205845, which occurs $6.7$ s after the \ac{GW} coalescence time.
The sky overlap integral is $\ln B_\mathrm{overlap}=1.35$ and visualized in \cref{fig:sky}.
{The false alarm rate is once every 2.2 years, which is consistent with the null hypothesis given that the total observation time of LIGO/Virgo is 443 days with at least two detectors running \citep{Nitz:2021zwj}.}
Since none of the search candidates meet the criteria that the false alarm rate is lower than 1-in-100 years, we conclude no statistically significant GW/long-GRB associations are found in our search.

We further retrieved any follow-up observations for the top-5 long GRB candidates in \cref{table} and noticed Swift/UVOT had detected an afterglow 109 s after the trigger time of GRB 170330A \citep{gcn20959}, the second top in our search results.
We are unaware of any additional follow-up observations associated with these long GRBs.
In addition, a visual inspection of the light curves of the GRBs in \cref{table} shows that GRB 170626A seems the only event that resembles the morphology of the light curve of GRB 211211A, i.e., with the main emission phase followed by some extended emission. Unfortunately, no firm conclusions can be drawn due to the lack of information, such as the GRB's host galaxy (thus, the redshift). Nevertheless, future improvements along this line can incorporate the GRB light curve properties into the ranking statistic as our knowledge evolves for long GRBs as the result of compact binary mergers.

Given the non-detection, we define an effective exclusion distance $d_\mathrm{ex}$ for each GW/long-GRB pair as the distance at which a 1.4-1.4 \msun (source frame mass) binary neutron star merger would have an \ac{SNR} of 8 at the GW trigger time while conservatively assuming the viewing angle is $30^\circ$. An \ac{SNR} of 8 is a conservative detection criteria for coincident gravitational-wave candidates.
 We marginalize the assumed source's location over the sky map given by Fermi-GBM/Swift-BAT and over the gravitational-wave polarization angle, assuming a uniform [0,2$\pi$] interval. 
The physical meaning of $d_\mathrm{ex}$ is the critical distance at which a source would have been conservatively detected, assuming it is indeed from a 1.4-1.4 \msun{} binary neutron star merger with sky map inferred from Fermi-GBM/Swift-BAT. 
For other component mass and inclination viewing angle assumptions, one can straightforwardly rescale $d_\mathrm{ex}$. 
The exclusion distance for all the GW/long-GRB search candidates is presented in the last column of \cref{table}. {A number of the GRBs in Table 1 were observed by Swift-BAT, which enabled us to further confirm the null results by examining the Swift-BAT catalog, which showed that none of those bursts are associated with any nearby galaxies within the $d_\mathrm{ex}$ reported in Table 1.}

\section{Discussion and Conclusion}

Inspired by the discovery of a kilonova located at 350 Mpc associated with a long GRB event \citep{Rastinejad:2022zbg}, this work performs the first multimessenger search for \ac{GW} from binary neutron star or neutron star-black hole mergers and potentially associated long GRB signals.
We set a 10 s time lag window for long GRB compared to a GW candidate and rank their correlation by their spatial overlap and the confidence of GW as real signals.
We present the complete search results in \cref{fig1} and \cref{table}, and do not find statistically significant associated pairs and thus conclude no detection.
With the null results, we report the exclusion distance as a threshold for the GW/GRB candidates, closer than that any 1.4-1.4 \msun{} binary neutron star mergers with an inclination angle 30$^\circ$ should have been detected through GW signals.

\setlength{\LTcapwidth}{\textwidth}
\begin{longtable*}[t]{cccccccccc}
\caption{The complete list of the search results.
We list the name of the \ac{GW} candidates, their chirp mass $\mathcal{M}_c = (m_1m_2)^{3/5}/(m_1+m_2)^{1/5}$, the name of the long \ac{GRB} candidates, their trigger time in UTC (coordinated universal time), the GW search statistic $\lambda_\mathrm{gw}$, the ranking statistic of GW/long-GRB association $\lambda_\mathrm{gw+grb}$, the inverse false alarm rate (IFAR), the GW observatories which record the data (H, L, and V stand for LIGO Hanford, LIGO Livingston and Virgo, respectively) and the exclusion distance $d_\mathrm{ex}$.
The nomenclature of GW candidates is in the form of YYMMDD$\_$HHMMSS which is the coalescence time in UTC.} \\
\hline 
&GW candidate 
&$\mathcal{M}_c / M_\odot$
&long GRB
&$t_\mathrm{GRB}$
&$\lambda_\mathrm{gw}$
&$\lambda_\mathrm{gw+grb}$
&IFAR/yr
&GWobs
&$d_\mathrm{ex}$/Mpc \\
\hline
1 & 200205\_201716 & 1.01 & GRB 200205845 & 2020-02-05 20:17:23.32 & 4.2 & 5.5 & 2.20 & HL & 281 \\
 2 & 170330\_222948 & 1.05 & GRB 170330A & 2017-03-30 22:29:51.34 & 6.9 & 6.7 & 1.99 & HL & 191 \\
 3 & 191213\_060532 & 1.01 & GRB 191213254 & 2019-12-13 06:05:33.02 & 4.1 & 4.1 & 0.91 & HLV & 286 \\
 4 & 151001\_082025 & 1.12 & GRB 151001348 & 2015-10-01 08:20:35.16 & 3.2 & 5.2 & 0.73 & HL & 156 \\
 5 & 190404\_070114 & 1.03 & GRB 190404293 & 2019-04-04 07:01:21.92 & 1.7 & 3.9 & 0.67 & HLV & 341 \\
 6 & 191110\_140525 & 1.20 & GRB 191110587 & 2019-11-10 14:05:34.99 & 0.3 & 3.2 & 0.54 & HLV & 231 \\
 7 & 191213\_040623 & 1.53 & GRB 191213A & 2019-12-13 04:06:23.92 & 2.7 & 1.8 & 0.29 & HLV & 363 \\
 8 & 190701\_094513 & 1.03 & GRB 190701A & 2019-07-01 09:45:20.83 & 3.4 & 1.9 & 0.27 & HLV & 295 \\
 9 & 170723\_161524 & 0.97 & GRB 170723677 & 2017-07-23 16:15:27.85 & 1.8 & 2.2 & 0.21 & HL & 176 \\
 10 & 170409\_024157 & 0.88 & GRB 170409112 & 2017-04-09 02:42:00.49 & 1.9 & 2.1 & 0.20 & HL & 194 \\
 11 & 190613\_040717 & 1.19 & GRB 190613A & 2019-06-13 04:07:18.31 & 1.7 & 0.6 & 0.16 & HLV & 343 \\
 12 & 200216\_090724 & 0.88 & GRB 200216380 & 2020-02-16 09:07:25.03 & 0.6 & -0.0 & 0.14 & HLV & 274 \\
 13 & 190827\_111245 & 1.57 & GRB 190827467 & 2019-08-27 11:12:48.54 & 1.3 & -0.5 & 0.11 & HL & 76 \\
 14 & 190508\_234118 & 1.24 & GRB 190508987 & 2019-05-08 23:41:24.14 & 0.6 & -0.5 & 0.11 & HLV & 389 \\
 15 & 190628\_123052 & 1.55 & GRB 190628521 & 2019-06-28 12:30:55.31 & 0.4 & -0.8 & 0.10 & HL & 313 \\
 16 & 170402\_065048 & 2.42 & GRB 170402285 & 2017-04-02 06:50:54.39 & 1.1 & -0.7 & 0.09 & HL & 184 \\
 17 & 190726\_152445 & 0.87 & GRB 190726642 & 2019-07-26 15:24:53.60 & 0.7 & -1.3 & 0.09 & HLV & 183 \\
 18 & 151029\_074936 & 1.40 & GRB 151029A & 2015-10-29 07:49:38.96 & 4.2 & -0.5 & 0.09 & HL & 65 \\
 19 & 191119\_061605 & 1.27 & GRB 191119261 & 2019-11-19 06:16:07.17 & 5.1 & -2.4 & 0.07 & HL & 192 \\
 20 & 170424\_101224 & 1.03 & GRB 170424425 & 2017-04-24 10:12:30.75 & 1.7 & -1.7 & 0.07 & HL & 129 \\
 21 & 190623\_110326 & 1.46 & GRB 190623461 & 2019-06-23 11:03:27.09 & 0.7 & -2.7 & 0.07 & HLV & 238 \\
 22 & 170626\_093721 & 1.17 & GRB 170626A & 2017-06-26 09:37:23.12 & 0.4 & -2.6 & 0.06 & HL & 137 \\
 23 & 200317\_004025 & 0.91 & GRB 200317028 & 2020-03-17 00:40:30.48 & 0.6 & -4.4 & 0.06 & HLV & 131 \\
 24 & 200117\_122400 & 1.05 & GRB 200117517 & 2020-01-17 12:24:06.53 & 2.9 & -4.9 & 0.06 & HLV & 213 \\
 25 & 190919\_181958 & 1.34 & GRB 190919764 & 2019-09-19 18:20:02.65 & 0.7 & -6.7 & 0.05 & HLV & 394 \\
 26 & 190805\_044554 & 1.84 & GRB 190805199 & 2019-08-05 04:46:00.97 & 3.2 & -10.8 & 0.04 & HLV & 204 \\
 27 & 190422\_160459 & 2.95 & GRB 190422670 & 2019-04-22 16:05:04.52 & 4.2 & -11.1 & 0.04 & HLV & 370 \\
 28 & 170825\_120003 & 1.12 & GRB 170825500 & 2017-08-25 12:00:05.99 & 0.2 & -6.7 & 0.04 & HLV & 186 \\
 29 & 170323\_012316 & 2.04 & GRB 170323058 & 2017-03-23 01:23:23.26 & 4.6 & -9.5 & 0.04 & HL & 121 \\
 30 & 190824\_144634 & 1.19 & GRB 190824A & 2019-08-24 14:46:39.57 & 0.8 & -inf & 0.04 & HLV & 292 \\
 31 & 151027\_224040 & 1.90 & GRB 151027B & 2015-10-27 22:40:40.66 & 7.4 & -inf & 0.03 & HL & 182 \\
 32 & 170629\_125329 & 2.17 & GRB 170629A & 2017-06-29 12:53:33.15 & 3.6 & -inf & 0.03 & HL & 195 \\
\hline 
\label{table}
\end{longtable*}

The observation reported by \cite{Rastinejad:2022zbg} opens up a promising future for a new type of GW multimessenger astronomy \citep[also see][]{2022arXiv220909256G}.
Richer structures of the burst, such as a precursor \citep{Xiao:2022quv} and an afterglow \citep{Mei:2022ncd, Zhang:2022fzj} are reported.
GW observation can provide irreplaceable information for the source properties of compact objects, including their mass, spin, tidal deformability, luminosity distance, and inclination angle, which is crucial to distinguishing different scenarios between binary neutron star, neutron star-black hole  \citep[e.g.,][]{Gompertz:2022jsg, Gao:2022iwn, Yang:2022qmy}.
Even with non-detection, GW can constrain the exclusion distance and provide evidence for alternative explanations for the engine of GRB 211211A, such as neutron star-white dwarf merger \cite{Yang:2022qmy}.
The next observation run of LIGO/Virgo/KAGRA is scheduled to start in March 2023 with further improved horizon distance.
Future (non-) detections will shed more light on the source property mystery of the GRB 211211A-like events.

We release all the data and scripts necessary to reproduce this work in \cite{wang_yi_fan_2022_7114357}.

\acknowledgements
YFW, AHN, and CDC acknowledge the Max Planck Gesellschaft and the Atlas cluster computing team at AEI Hannover for support.  
BBZ acknowledges the support by the National Key Research and Development Programs of China (2018YFA0404204), the National Natural Science Foundation of China (Grant Nos. 11833003, U2038105, 12121003), the science research grants from the China Manned Space Project with NO.CMS-CSST-2021-B11, and the Program for Innovative Talents, Entrepreneur in Jiangsu.
This research has made use of data, software and/or web tools obtained from the Gravitational Wave Open Science Center (https://www.gw-openscience.org), a service of LIGO Laboratory, the LIGO Scientific Collaboration and the Virgo Collaboration. LIGO is funded by the U.S. National Science Foundation. Virgo is funded by the French Centre National de Recherche Scientifique (CNRS), the Italian Istituto Nazionale della Fisica Nucleare (INFN) and the Dutch Nikhef, with contributions by Polish and Hungarian institutes. We acknowledge the use of public data from the Fermi Science Support Center (FSSC) and the Swift data archive.

\bibliography{reference}
\end{CJK*}
\end{document}